\documentclass{pasa}

\jid{PASA}
\doi{10.1017/pas.\the\year.xxx}
\jyear{2013}

\usepackage{amsmath}
\usepackage{txfonts}

\def\degr{\hbox{$^\circ$}}

\newcommand{\keyi}[2]{\mbox{{\tt #1\hspace{1pt}}{$#2$}\/}}
\newcommand{\CTYPE}[1]{\keyi{CTYPE}{#1}}

\newcommand{\keyv}[1]{\mbox{{\tt #1}}}

\DeclareMathOperator{\fmod}{fmod}
\DeclareMathOperator{\sgn}{sgn}

\title[Representing the {\it ``butterfly''} projection in FITS]{Representing
the {\it ``butterfly''} projection in FITS -- \\ projection code \keyv{XPH}}

\author[Calabretta and Lowe]
  {Mark R. Calabretta
   \thanks{Email: mcalabre@atnf.csiro.au}\\
   \affil{CSIRO Astronomy and Space Science,
          PO Box 76, Epping, NSW 1710, Australia}
   \and Stuart R. Lowe\\
   \affil{Las Cumbres Observatory Global Telescope Network,
          6740 Cortona Drive Suite 102, Goleta, CA 93117, US}}

\begin{document}

%==================================================================== Abstract

\begin{abstract}
The {\it ``butterfly''} projection is constructed as the polar layout of the
HEALPix projection with $(H,K) = (4,3)$.  This short article formalises its
representation in FITS.
\end{abstract}

%==================================================================== Keywords

\begin{keywords}
  methods: data analysis --
  methods: statistical --
  techniques: image processing --
  astronomical data bases: miscellaneous --
  cosmic background radiation --
  cosmology: observations
\end{keywords}

\maketitle

%================================================================ Introduction

\section{INTRODUCTION}
\label{sec:intro}

\begin{figure}
  \centering
  \includegraphics[width=170pt]{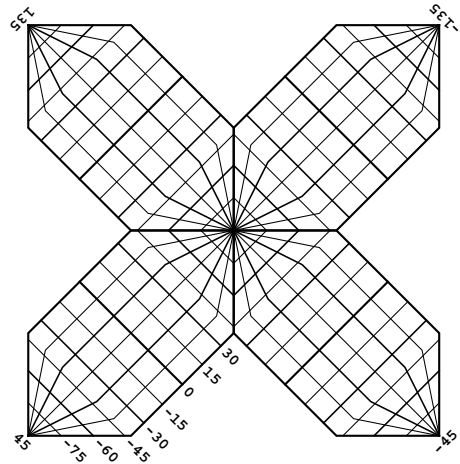}
  \caption[]{The {\it butterfly} projection at the same scale as the
    graticules depicted in WCS Paper II.}
  \label{fig:XPH}
\end{figure}

The {\it butterfly} projection was described briefly in Section~3 of
Calabretta \& Roukema \shortcite{MHG} as the polar variant of the HEALPix
(\keyv{HPX}) projection with $(H,K) = (4,3)$.  It is constructed by splitting
the latter into four {\em gores} along the $0\degr$ and $\pm 90\degr$
meridians, rotating them by $45\degr$ plus appropriate multiples of $90\degr$,
and joining them at the pole to produce an $\times$ or `butterfly'-shaped
layout as depicted in Figure~\ref{fig:XPH}.  This achieves a 75\% filling
factor of the enclosing square which compares favourably to 48\% for the
rotated \keyv{HPX} projection.

The enhanced filling factor, together with the fact that the required
$45\degr$ rotation of the gores is a natural part of the projection, suggest
that the butterfly projection is more suitable for storing HEALPix
single-pixelisation data than the rotated \keyv{HPX} projection which itself
is better suited to the double-pixelisation \cite{MHG}.

The purpose of this article is to formalise the representation of the
butterfly projection in FITS \cite{FITS}.

%================================================================= XPH in FITS

\section{\keyv{XPH} IN FITS}
\label{sec:XPH}

The butterfly projection will be denoted in FITS with algorithm code
\keyv{XPH} in the \CTYPE{ia} keywords for the celestial axes.  As it is
constructed with the pole of the native coordinate system at the reference
point, we set
\begin{equation}
   (\phi_0, \theta_0)_{\keyv{XPH}} = (0,90\degr).
\end{equation}
The \keyv{XPH} projection is constructed by rearranging the gores of the
\keyv{HPX} projection, whereby the scale at the reference point is inherited
from the scale at the poles of the \keyv{HPX} projection.  In fact, the scale
varies with direction (is non-conformal), with circles projected as squares.
Thus, the \keyv{XPH} projection is not scaled true at the reference point in
the sense discussed in Section~5 of WCS Paper II \cite{WCS2}.  The map scale
is expanded by $\pi\sqrt{3}/4$ ($\approx 1.36$) in the $x$ and $\vary$
directions, reducing to $\pi\sqrt{3/2}/4$ ($\approx 0.96$) along the
diagonals.

%-------------------------------------------------------------------- floating

%================================================ Forward projection equations

\subsection{Projection equations}

The projection equations for \keyv{XPH}, together with their inverses,
expressed in degrees as required by FITS, are as follows.

Assuming that the native longitude, $\phi$, is normalised in the range
$[-180\degr, 180\degr)$, then
\begin{align}
  (x,\vary) & = \left\{
                  \begin{array}{l@{\hskip 8pt}l@{\hskip 3pt}r@{\hskip 3pt}l@{\hskip 2pt}r}
                    \kappa \, (-\xi'+\eta',-\xi'-\eta') &
                      \ldots & -180\degr & \le \phi < & -90\degr , \\
                    \kappa \, (\hphantom{+}\xi'+\eta',-\xi'+\eta') &
                      \ldots &  -90\degr & \le \phi < &   0\degr , \\
                    \kappa \, (\hphantom{+}\xi'-\eta',+\xi'+\eta') &
                      \ldots &    0\degr & \le \phi < &  90\degr , \\
                    \kappa \, (-\xi'-\eta',+\xi'-\eta') &
                      \ldots &   90\degr & \le \phi < & 180\degr ,
                  \end{array}
                \right. \label{eq:xy} \\
  (\xi',\eta') & = (\xi-45\degr, \eta-90\degr) .
\end{align}
where $\kappa = \sqrt{2}/2$.  The native latitude that divides the equatorial
and polar zones is $\theta_\times = \sin^{-1}(2/3)$.  In the equatorial zone,
$|\,\theta\,| \le \theta_\times$, we have
\begin{align}
     \xi & = \psi , \\
    \eta & = \frac{135\degr}{2} \sin\theta ,
\end{align}
and in the polar zones, where $|\,\theta\,| > \theta_\times$
\begin{align}
     \xi & = 45\degr + (\psi - 45\degr) \,\sigma , \\
    \eta & = \sgn\theta \, ( 90\degr - 45\degr \sigma ) ,
\end{align}
where $\sgn\theta$ gives the algebraic sign of $\theta$, and
\begin{align}
    \psi & = (\phi + 180\degr) \pmod{90\degr} , \label{eq:psi} \\
  \sigma & = \sqrt{3 (1 - |\sin\theta\,|\,)} .  \label{eq:sigma}
\end{align}

%================================================ Inverse projection equations

\subsection{Deprojection equations}

To invert the projection equations first compute $(\xi,\eta)$ via
\begin{align}
  (\xi, \eta)  & = (\xi'+45\degr, \eta'+90\degr) , \\
  (\xi',\eta') & = \left\{
                     \begin{array}{l@{\hskip 8pt}l@{\hskip 3pt}l}
                       \kappa \, (-x-\vary,\hphantom{+}x-\vary) &
                         \ldots & x \le 0, \vary > 0 , \\
                       \kappa \, (\hphantom{+}x-\vary,\hphantom{+}x+\vary) &
                         \ldots & x < 0, \vary \le 0 , \\
                       \kappa \, (\hphantom{+}x+\vary,-x+\vary) &
                         \ldots & x \ge 0, \vary < 0 , \\
                       \kappa \, (-x+\vary,-x-\vary) &
                         \ldots & x > 0, \vary \ge 0 ,
                     \end{array}
                   \right.
\end{align}
where $\kappa = \sqrt{2}/2$ as before.  Then
\begin{align}
  \phi & = \left\{
                  \begin{array}{l@{\hskip 8pt}l@{\hskip 3pt}l}
                    \psi - 180\degr &
                      \ldots & x \le 0, \vary > 0 , \\
                    \psi - 90\degr &
                      \ldots & x < 0, \vary \le 0 , \\
                    \psi &
                      \ldots & x \ge 0, \vary < 0 , \\
                    \psi + 90\degr&
                      \ldots & x > 0, \vary \ge 0 .
                  \end{array}
                \right.
\end{align}
In the equatorial zone where $|\,\eta\,| \le 45\degr$
\begin{align}
    \psi & = \xi , \\
  \theta & = \sin^{-1} \left( \frac{2 \eta}{135\degr} \right) ,
\end{align}
and in the polar zones, where $|\,\eta\,| > 45\degr$
\begin{align}
    \psi & = 45\degr + (\xi - 45\degr) / \sigma , \\
  \theta & = \sgn\eta \sin^{-1} \left( 1 - \frac{\sigma^2}{3} \right) ,
    \label{eq:theta}
\end{align}
where
\begin{align}
    \sigma & = \frac{90\degr - |\,\eta\,|}{45\degr} .
\end{align}

%======================================================== Implementation notes

\section{IMPLEMENTATION NOTES}

This section records some problems that may arise in implementing these
equations at the highest levels of numerical precision.

%================================================================ Equation (8)

\subsection{Equation~(\ref{eq:psi})}

A subtle problem arises from the innocent looking Equation~(\ref{eq:psi}) when
$\phi < 0$ but very close to zero; in this case $\psi$ should be just slightly
less than $90\degr$.  However, due to the loss of numerical precision that
results from adding $180\degr$ to $\phi$, or simply from taking the modulo
$90\degr$ (e.g.\ via the $\fmod$ function in C), application of
Equation~(\ref{eq:psi}) may instead yield $\psi = 0$, effectively as though
$\phi = 0$.  After computing $(\xi',\eta')$, the problem then arises by
selecting the $\phi < 0$ option in Equation~(\ref{eq:xy}) when in fact the
$\phi = 0$ option would be the appropriate one in this case.

A simple solution is, when computing $\psi$, also to recompute $\phi$ as
$(\phi + 180\degr) - 180\degr$ as this will apply the same numerical rounding
to $\phi$ as occurred in computing $\psi$, thereby ensuring selection of the
appropriate option in Equation~(\ref{eq:xy}).

%====================================================== Equations (9) and (16)

\subsection{Equations~(\ref{eq:sigma}) and (\ref{eq:theta})}

Numerical precision may be lost in Equations~(\ref{eq:sigma}) and
(\ref{eq:theta}) when $\sigma$ is very close to zero which unfortunately
occurs in the neighbourhood of the reference point.  This precision may be
recovered by rewriting Equation~(\ref{eq:theta}) using the small angle
trigonometric formulae which allows the expression in $\sigma^2$ to be
replaced by one in $\sigma$.  Thus
\begin{align}
  \theta & = 90\degr - \frac{180\degr}{\pi} \sqrt{\frac{2}{3}} \sigma
  \label{eq:theta2}
\end{align}
which is applicable for $\sigma < 10^{-4}$ for calculations in double
precision IEEE floating point.  Equation~(\ref{eq:sigma}) may then be replaced
by the inverse of this
\begin{align}
  \sigma & = (90\degr - \theta) \frac{\pi}{180\degr} \sqrt{\frac{3}{2}}
\end{align}
which is applicable for $\theta$ greater than the value obtained from
Equation~(\ref{eq:theta2}) with $\sigma = 10^{-4}$.

%================================================================== Conclusion

\section{CONCLUSION}

\keyv{XPH} has been implemented in version 4.18 and later versions of
\keyv{WCSLIB} \cite{WCSLIB} which is distributed under the Lesser GNU
General Public License (LGPL).

As of version 4.3, \keyv{WCSLIB} has included a utility program that converts
1D HEALPix pixelization data stored in a variety of forms in FITS, including
ring or nested organization in a binary table extension, into a 2D primary
image array optionally with \keyv{HPX} or \keyv{XPH} coordinate
representation.

%============================================================ Acknowledgements

\begin{acknowledgements}
We wish to thank Dr.\ Paddy Leahy of the Jodrell Bank Centre for Astrophysics
for motivating this work.

The Australia Telescope is funded by the Commonwealth of Australia for
operation as a National Facility managed by CSIRO\@.
\end{acknowledgements}

%================================================================ Bibliography

\end{document}